\documentstyle[12pt,fleqn]{article}
\begin{document}
\baselineskip 7mm

\begin{center}
Physical Review Letters 74 (1995) 4571; 76 (1996) 2205 (erratum)
\vfill

{\Large Nonlocal effects in Fock space}
\\[1cm] Asher Peres$^*$\bigskip

{\sl Department of Physics,
Technion---Israel Institute of Technology,\\ 32 000 Haifa, Israel}\vfill

25 January 1995\\[1cm]

{\bf Abstract}
\end{center}

If a physical system contains a single particle, and if two distant
detectors test the presence of linear superpositions of one-particle and
vacuum states, a violation of classical locality can occur. It is due to
the creation of a two-particle component by the detecting process
itself.\vfill

\vfill

\noindent $^*$ Electronic address: peres@photon.technion.ac.il\newpage

It has been known for a long time that quantum systems consisting of two
[1] or more [2, 3] distant particles display remarkable nonlocal effects.
Recently, a similar nonlocal effect was predicted by Hardy [4] for a
quantum system involving no more than one photon. Hardy considered a
state
\begin{equation}
|\Psi\rangle=p\,{|1\rangle}_a\,{|0\rangle}_b+
	     q\,{|0\rangle}_a\,{|1\rangle}_b+
	     r\,{|0\rangle}_a\,{|0\rangle}_b\,,
\end{equation}
and various tests performed by two distant observers, Alice and Bob, who
find a violation of classical locality. In the above equation,
${|0\rangle}_a$  and ${|1\rangle}_a$ denote the vacuum and one-particle
states in a beam directed toward Alice; ${|0\rangle}_b$ and
${|1\rangle}_b$ likewise refer to a beam directed toward Bob; and $p,
\,q, \,r$ are numerical coefficients, none of which is zero. Hardy gave
explicit instructions on how to actually perform these experiments, by
means of beam splitters and parametric down conversion processes. The
abundance of technical details is helpful for convincing the reader that
the experiment is indeed feasible, but it somewhat obscures the
origin of the nonlocality.

As shown below, the latter is simply due to the creation of a component
${|1\rangle}_a{|1\rangle}_b$ by the detecting process itself. To
simplify the discussion, I shall restrict it to the case where
\begin{equation}
|\Psi\rangle=(\,{|1\rangle}_a\,{|0\rangle}_b-
	     {|0\rangle}_a\,{|1\rangle}_b\,)/\sqrt{2},
\end{equation}
is a pure one-particle state, without vacuum component. Such a state
could also be written without invoking Fock space notations, since it
involves only ordinary quantum mechanics, with a given number of
particles (one). However, it is impossible to repeat Hardy's argument by
using a first-quantized formalism, for reasons that will soon be clear.

Note that the right hand side of Eq.~(2) has the same structure as the
singlet state of a {\em pair\/} of particles of spin~$1\over2$, if we
reinterpret $|0\rangle_a$ and $|1\rangle_a$ as representing
particle $a$ with spin up and down, respectively, and likewise for
particle $b$. The route to nonlocality is now obvious.

Both Alice and Bob have a choice of two different experiments.
One is to test the mere presence of a particle, by measuring the
projection operators $P_a$ and $P_b$ on the one-particle states
$|1\rangle_a$ and $|1\rangle_b$, respectively. Alice can also
opt to test the projection operator $P_{a'}$ on the state
${1\over2}(\,{|1\rangle_a}+\sqrt{3}{|0\rangle_a})$, namely a coherent
superposition of one-particle and vacuum states. Independently of her
decision, Bob can choose to test $P_{b'}$, the projection operator on
${1\over2}(\,{|1\rangle_b}-\sqrt{3}{|0\rangle_b})$. There are therefore
four different experiments, and quantum theory makes the following
predictions, for the state $|\Psi\rangle$ in Eq.~(2):
\begin{equation}
\langle P_{a'}\rangle=\langle P_{b'}\rangle=0.5, \end{equation}
\begin{equation}
\langle P_a\,P_b\rangle=0, \end{equation}
\begin{equation}
\langle P_a\,P_{b'}\rangle=\langle P_{a'}\,P_b\rangle=
\langle P_{a'}\,P_{b'}\rangle=0.375. \end{equation}
These results violate the Clauser-Horne inequality [5] (a variant of
Bell's inequality), namely
\begin{equation}
  0\leq\langle P_{a'}+P_{b'}-P_{a'}\,P_{b'}-P_{a'}\,P_b
  -P_a\,P_{b'}+P_a\,P_b\rangle\leq1.
\end{equation}
For the given $|\Psi\rangle$, the actual value of the above expression
is $-0.125$.

Obviously, the total number of particles is not conserved when we
measure $P_{a'}$ or $P_{b'}$, since these operators do not commute with
the number operator. The apparatuses used by Alice and Bob must be able
to create new particles, or supply some of their own.  Nonlocal effects
may thus appear for an initial state that contains a single particle,
provided that the final state may contain two. (I did not include in
this discussion the numerous auxiliary particles in the two measuring
apparatuses, as Hardy did in ref.~[4], because quantum mechanical
probabilities do not depend on the detailed structure of these
apparatuses, and it is both customary [1--3] and legitimate [6] to
ignore the latter.)\bigskip

This work was supported by the Gerard Swope Fund, and the Fund for
Encouragement of Research.\bigskip

\begin{enumerate}
\item J. S. Bell, Physics {\bf 1}, 195 (1964).
\item D. M. Greenberger, M. Horne, and A. Zeilinger, in {\sl Bell's
Theorem, Quantum Theory, and Conceptions of the Universe\/}, ed. by M.
Kafatos, Kluwer, Dordrecht (1989) p.~69.
\item N. D. Mermin, Am.\ J.\ Phys. {\bf 58}, 731 (1990).
\item L. Hardy, Phys.\ Rev.\ Letters {\bf 73}, 2279 (1994).
\item J. F. Clauser and M. A. Horne, Phys.\ Rev.\ D {\bf 10}, 526
(1974).
\item A. Peres, {\sl Quantum Theory: Concepts and Methods\/}, Kluwer,
Dordrecht (1993), Chapt.~12.
\end{enumerate}
\end{document}